\documentclass[submission, Phys]{SciPost}
\begin{document}

\begin{center}{\Large \textbf{
Numerical Simulation of the Coupling between Split-Ring Resonators and Antiferromagnetic Magnons
}}\end{center}

% TODO: write the author list here. Use initials + surname format.
% Separate subsequent authors by a comma, omit comma at the end of the list.
% Mark the corresponding author with a superscript *.
\begin{center}
Daniel M. Heligman\textsuperscript{1*},
R. Vald\'{e}s Aguilar\textsuperscript{1$\dagger$}
\end{center}

\begin{center}
{\bf 1} Department of Physics, The Ohio State University, Columbus, OH, USA
\\
* heligman.1@osu.edu\\
$\dagger$ valdesaguilar.1@osu.edu
\end{center}

\begin{center}
\today
\end{center}

% For convenience during refereeing: line numbers
%\linenumbers

\section*{Abstract}
{\bf
% TODO: write your abstract here.
We report on the results of simulations of the terahertz response of a split ring resonator (SRR) metamaterial coupled to a hypothetical antiferromagnetic material characterized by a magnon resonance. The simulations were done using finite difference time domain (FDTD) techniques. By adjusting the magnon frequency we find a hybridization of the resonant normal modes of the SRR and the magnon manifested as an avoided crossing. By varying the physical separation between the metamaterial and the antiferromagnet with a dielectric spacer, we evaluated the coupling strength between the two.
}


%\vspace{10pt}
%\noindent\rule{\textwidth}{1pt}
%\tableofcontents\thispagestyle{fancy}
%\noindent\rule{\textwidth}{1pt}
%\vspace{10pt}

\section{Introduction}
Metamaterials are artificial structures where a unit (a meta-atom) is repeated in space and that exhibit properties that are analogous to naturally occurring materials \cite{veselago1968electrodynamics,pendry1999magnetism}. When the meta-atom dimensions are smaller than the radiating wavelength, the metamaterial will behave like a smooth continuous medium. Since one has the freedom to manipulate the dimensions of meta-atom, one can create properties not observed in naturally occurring materials as well. The structure can be tweaked to create effects such as negative index of refraction, perfect lensing, and giant optical activity \cite{pendry2000negative,shelby2001experimental,padilla2006negative,pendry2006controlling,shalaev2007optical,soukoulis2007negative}. Like naturally occurring materials, metamaterials can respond to electromagnetic radiation resonantly at certain frequencies. The existence of these resonant modes suggest that metamaterials can couple to the normal modes of other materials, similar to how modes couple in naturally occurring matter. Coupling in normal materials can manifest in the form of polaritons, when light couples to matter, electromagnons, when magnons (spin waves) couple to phonons, or electron-phonon coupling in superconducting materials \cite{mills1974polaritons,sushkov2007electromagnons,schrieffer2018theory,tinkham2004introduction,PhysRev.108.1175}. 

\begin{figure}[!t]
    \centering
    \includegraphics[width=.9\columnwidth]{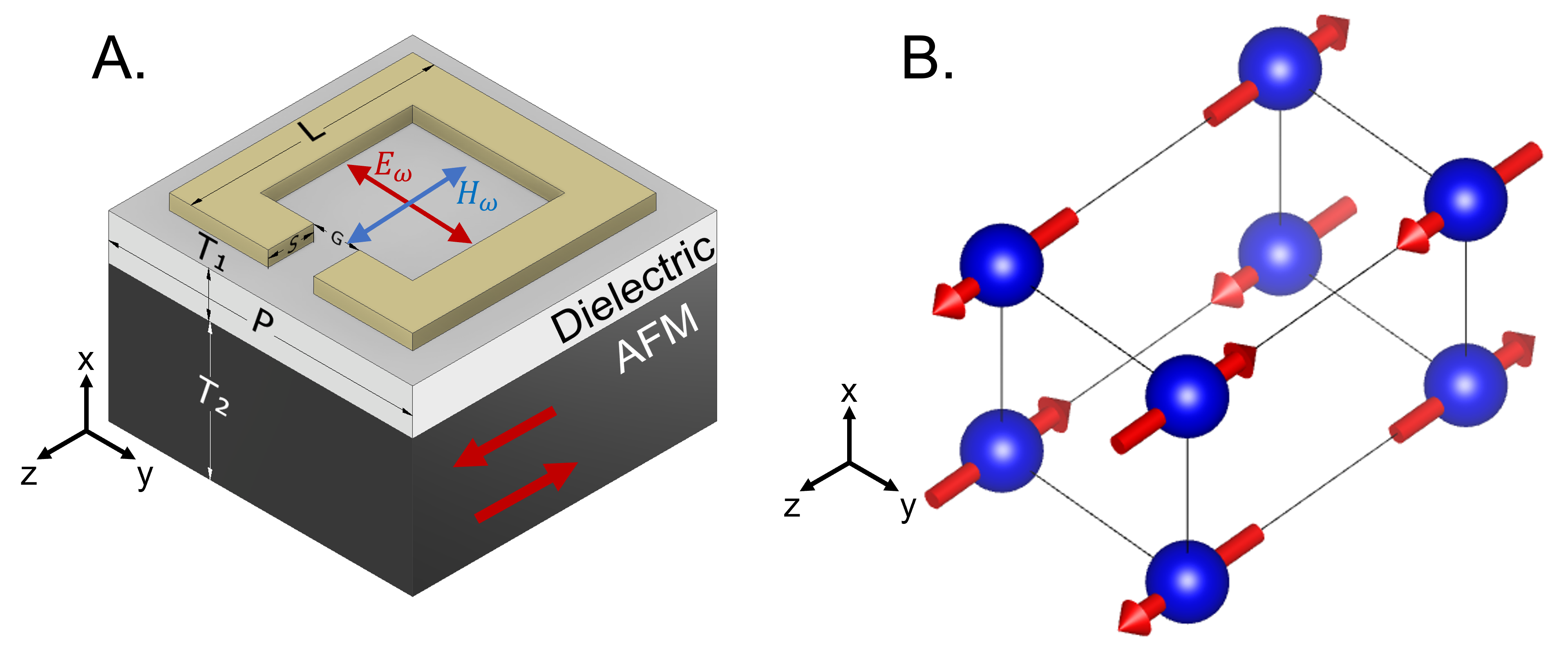}
    \caption{\textbf{A.} Geometry used for the metamaterial/AFM material structure. L, S, G, T$_1$, T$_2$, and P represent the side length of the metamaterial, the strip width, the gap length, thickness of the spacer, the thickness of the magnetic material, the thickness of the substrate, and the periodicity, respectively. For this set of simulations L = 30 $\mu$m, S = 5 $\mu$m, G = 5 $\mu$m, T$_1$ ranges between 0 $\mu$m and 3 $\mu$m, T$_2$ = 500 $\mu$m, and P = 35 $\mu$m, respectively. The polarization of the electric field is parallel to the gap of the SRR along the y-axis. \textbf{B.} A diagram depicting the simulated magnetic structure. It represents an antiferromagnetic crystal with 4-fold rotational symmetry along the z-axis with an easy-axis pointing in the z-direction. The spins are parallel to the H-field to prevent excitation from the THz pulse.}
    \label{fig:geometry}
\end{figure}

Similar to how naturally occurring materials can contain different types of modes that couple to one another, certain groups of metamaterials have shown coupling effects between different components of the meta-atom's structures. One such example is coupling light and dark modes. The former arises from an element of the structure that is directly coupled to the external radiation, while the latter manifests in a second element that is exclusively driven by the first element. This combined system will exhibit a hybrid mode in the form of a Fano resonance \cite{luk2010fano,singh2009coupling}. One consequence of this coupling effect that has been extensively studied is artificially generated electromagnetically induced transparency, in which a portion of an absorption line becomes more transmissive with increasing coupling \cite{tassin2009low,liu2009plasmonic,papasimakis2008metamaterial}. 

There has also been an effort in researching the possible interactions of metamaterials with naturally occurring materials. One such example has been observed in ref.\cite{shelton2011strong} in which a split ring resonator (SRR) metamaterial was coupled to a thin film of SiO$_2$. It was found that when the metamaterial mode was frequency matched with that of an infrared active phonon, a Rabi splitting occurred due to the near-field interaction between the SRR resonance mode and the SiO$_2$ phonon. Other studies have also explored the electrical coupling between natural materials and metamaterials \cite{campione2014electrodynamic,benz2015control,grebenchukov2020asymmetric,gabbay2011interaction, schalch2019strong, scalari2012ultrastrong}. In one instance, the possibility of utilizing metamaterials to excite a magnetic response has been reported \cite{mukai2014antiferromagnetic}. They showed that the magnetic field from a THz pulse could be amplified 20-fold with the incorporation of SRRs. This allowed for the measurement of the antiferromagnetic resonance. While this report focused on the near-field interaction between metamaterials and antiferromagnets, not much has been explored in how the coupling can affect the far field response. Since metamaterials can be pictured as microcavities, knowing how these two materials interact could inspire further research such as in cavity spintronics \cite{benz2015control,harder2018cavity}. The effects of coupling an antiferromagnet (AFM) with SRRs on the THz response was numerically evaluated in this report by performing finite difference time domain simulations (FDTD) of a THz pulse propagating through a metamaterial/AFM structure. 

\begin{figure}
    \centering
    \includegraphics[width = 0.5\columnwidth]{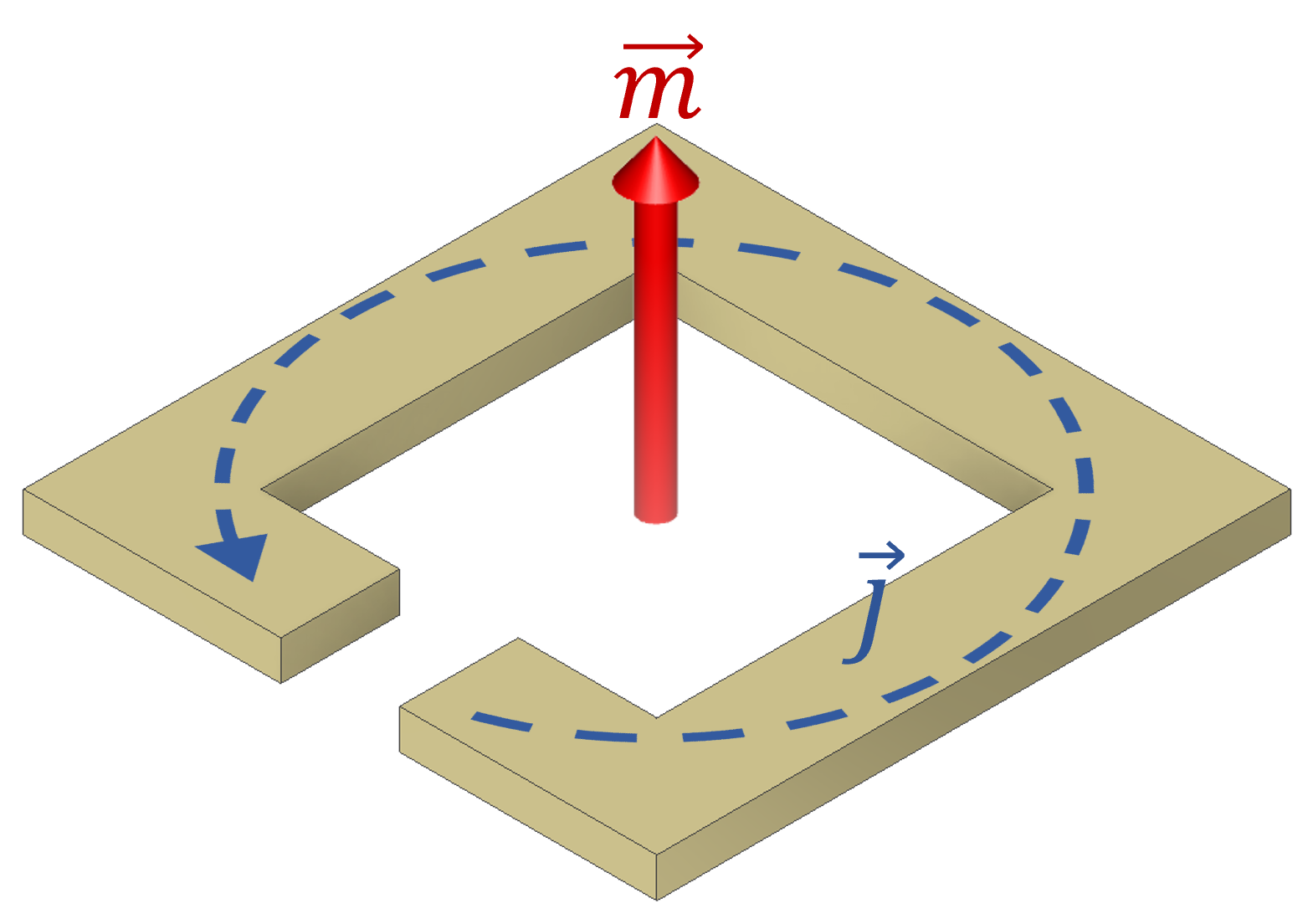}
    \caption[]{Magnetic moment of LC resonance of an SRR. For this mode, the H-fields manifest as a magnetic dipole moment, $\overrightarrow{m}$, oriented perpendicular to the meta-surface and through the ring of the SRR, and this is generated by a surface current $\overrightarrow{J}$ circulating around the SRR.}
    \label{fig:srrnearfields}
\end{figure}

For this report, the transmission of an SRR \cite{o2002magnetic} coupled to an antiferromagnetic substrate was studied. This particular AFM emulates a hypothetical two-sublattice easy-axis antiferromagnet with a 4-fold symmetry along the z-axis. With the easy-axis pointing in the z-direction, the spins are consequently aligned and anti-aligned in the z-direction (fig. \ref{fig:geometry}). This type of AFM would exhibit resonant modes in the form of magnons, or spin waves, in the region between microwaves and terahertz frequencies \cite{rezende2019introduction}. To evaluate the coupling between the two materials, the magnon resonant frequency of the AFM was tuned so that its interaction with the SRRs can be analyzed. The frequency was specifically tuned around the LC mode of the SRR. This particular mode, which is named for its quasi-static resemblance to the normal mode of an LC circuit, shows a strong uniform magnetic dipole moment oriented perpendicular to the meta-surface through the ring of the SRR (Fig. \ref{fig:srrnearfields}) \cite{Padilla2006magnetic}. By setting the THz polarization such that the magnetic field of the THz pulse is parallel to the spin direction, the magnon mode can only be excited by the SRR magnetic dipole moment pointing along the x-direction. The strength of the magnetic response and the separation between the SRRs and AFM was varied as to better understand the coupling properties between these two modes and their coupling strength. We have found that the coupling occurs best in materials with high permeability and that the coupling remains relatively strong when the SRRs and AFM are within a few $\mu$m from one another. These coupling effects manifest as the hybridization of the SSR/AFM fundamental modes in the form of an Rabi splitting at the LC-resonance of the SRRs. At high frequencies, a persistent hybrid magnon mode is observed, allowing for the observation of a selection rule forbidden mode that is not excitable in the configuration shown in fig.\ref{fig:geometry}A when there is no metamaterial present.

\section{Methods}

To perform our simulations, we utilized FDTD numerical techniques using MIT's Electromagnetic Equation Propagation (MEEP) software \cite{oskooi2010meep}. The specific geometry we simulated is depicted in Fig. \ref{fig:geometry}A which consists of a stack of different materials: SRR, dielectric spacer (for evaluating the extent of the coupling), and the AFM. To simulate an array of this structure, a periodic boundary condition was defined for the computational cell walls in the z and y directions. The propagation direction of the electromagnetic wave is the x-axis, and the electric (magnetic) field of the THz pulse, $E_\omega$ ($H_\omega$), points in the y- (z-) axis.

We have chosen to use the SRR geometry for the metamaterial, as this shape is simple and intuitive, and exhibits strong confinement of magnetic near-fields near its fundamental LC resonance. For the SRR, we chose the dimensions for L, S, G, P, and T$_2$ to be 30 $\mu$m, 5 $\mu$m, 5 $\mu$m, 35 $\mu$m and 500 $\mu$m, respectively. These values allow for the modes of the SRRs to be within the range of resonant frequencies that were chosen for the AFM. As the resistance of metal in the terahertz regime is low, the SRR's material was set to be a perfect electric conductor to reduce computational cost. 

AFM materials generally contain a magnon mode that emerges and increases in frequency as temperature decreases below the magnetic ordering temperature, T$_{N\acute{e}el}$, as the spins become ordered. For this simulation it is natural to keep the dimensions of the SRR fixed and shift the magnon frequency of the AFM to emulate the temperature dependence of its frequency. The AFM was modeled with a magnetic permeability whose frequency dependence is defined by the Lorentz-Drude model,

\begin{equation}
    \mu_r(f) = 1 + \frac{\sigma f_0^2}{f_0^2-f^2-if\gamma_0},
    \label{eq:lorentz}
\end{equation}
where $\sigma$ is the oscillator strength which quantifies the strength of coupling of the AFM to magnetic fields, $f_0$ is the magnon frequency, and $\gamma_0$ is the damping constant of the magnon, inversely related to its lifetime.

A range of fundamental frequencies, $f_0$, from 100 GHz to 2 THz at 50 GHz increments was tested. The oscillator strength, $\sigma$, for each $f_0$ analyzed was set to a particular value that preserves the maximum of the imaginary part of permeability. To analyze the effects that oscillator strength has on the coupling, it was desired to evaluate a wide range of different oscillator strengths representing AFMs with the weakest permeabilities and beyond. A baseline value of $\sigma$ ($\sigma_0$) was chosen such that the maximum value of the imaginary part of permeability is 0.03. To test different oscillator strengths, $\sigma_0$ was multiplied by a prefactor of $\sigma/\sigma_0$, with this value ranging between 1 and 50 at increments of 5. For all frequencies, the damping coefficient, $\gamma_0$, was chosen to be fixed at a value of 66 GHz as it corresponds to experimental values from materials such as CaFe$_2$O$_4$ \cite{mai2019optical}. Along with the magnetic properties, we have chosen the AFM material's relative permittivity to be 6, which is a typical value for an insulating material. Since the spins of highly ordered easy-axis antiferromagnetic materials tend to align in one direction and magnon excitations occur when magnetic fields are perpendicular to spin direction, the material was simulated by an anisotropic permeability tensor. This tensor simulates a material with spins oriented in the z-direction to prevent excitation from external THz magnetic fields (Fig. \ref{fig:geometry}). The result is a tensor, whose $\mu_{zz}$ is unity and $\mu_{xx}=\mu_{yy}$ components are dispersive \cite{rezende2019introduction,safin2020excitation,kaganov1997magnons},
\begin{equation}
   \stackrel{\leftrightarrow}{\mu}_r(f) = 
   \begin{pmatrix}
    \mu_r(f) & 0 & 0\\
    0 & \mu_(f) & 0\\
    0 & 0 & 1
    \end{pmatrix},
    \label{eq:permeability}
\end{equation}
where $\mu_r(f)$ is defined in eq. \ref{eq:lorentz}.

A pulse of THz radiation was simulated passing through the SRR/AFM system. The resulting signal was referenced to the signal that passed through a non-magnetic material whose permittivity matches that of the AFM. This ensured that the phase difference between the sample signal and reference signal was minimal. These two signals were Fourier transformed and the ratio between them was taken to derive the transmission. The coupling effects observed in the transmission were then analyzed. To evaluate the extent of the coupling, the distance between the SRRs and AFM was varied. This was done by inserting a dielectric spacer of thickness T$_1$ between the SRRs and AFM. For the simulations, we set the permittivity of the substrate to match that of the AFM to maximize the distance that the near fields penetrate. To test the coupling strength, T$_1$ was varied between 0 $\mu$m and 3 $\mu$m at 0.5 $\mu$m increments. The coupling constant for each thickness of each spacer material was then calculated.

\begin{figure*}[!t]
    \begin{center}
    \centering
    \includegraphics[width=0.48\columnwidth]{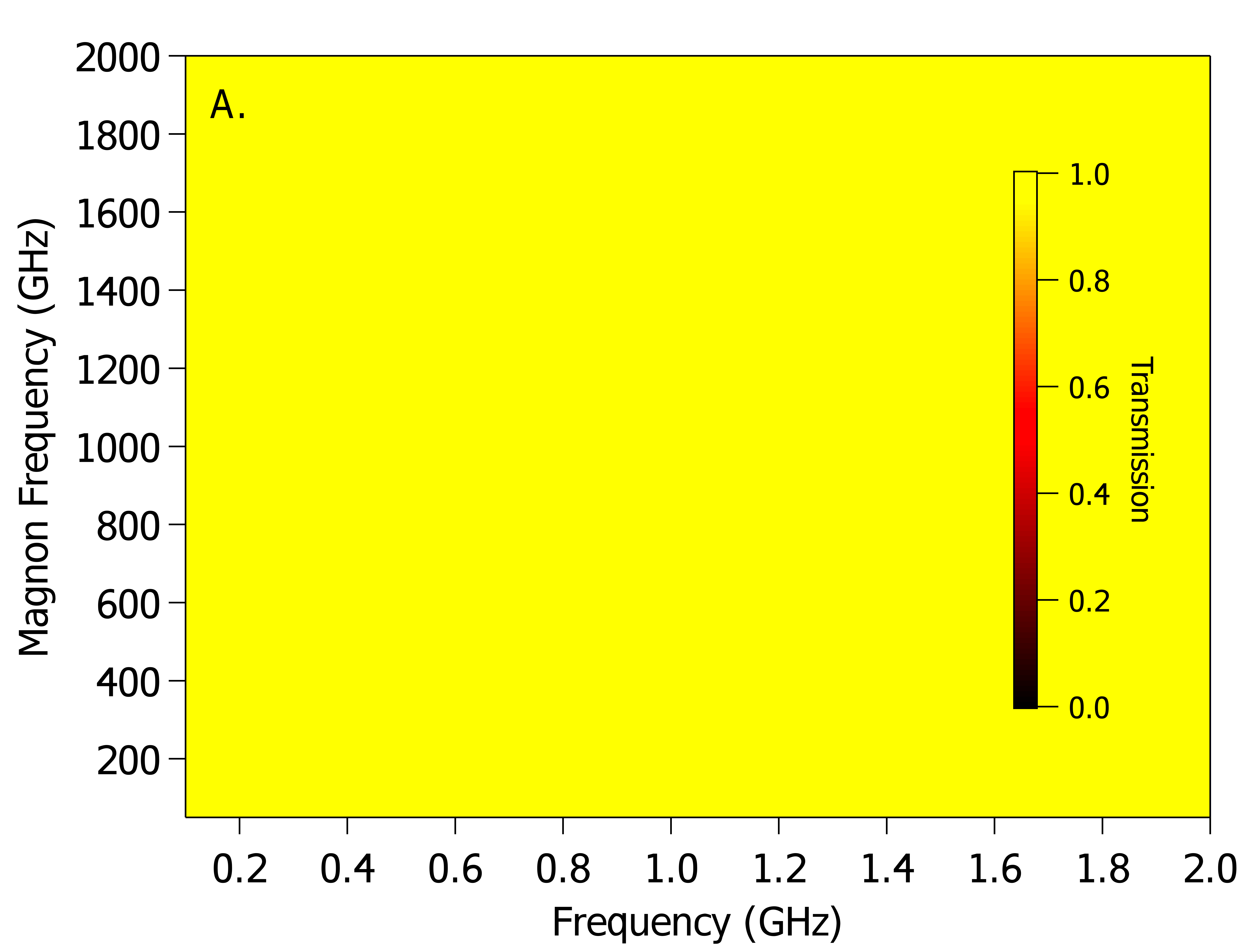}
     \centering
    \includegraphics[width=0.48\columnwidth]{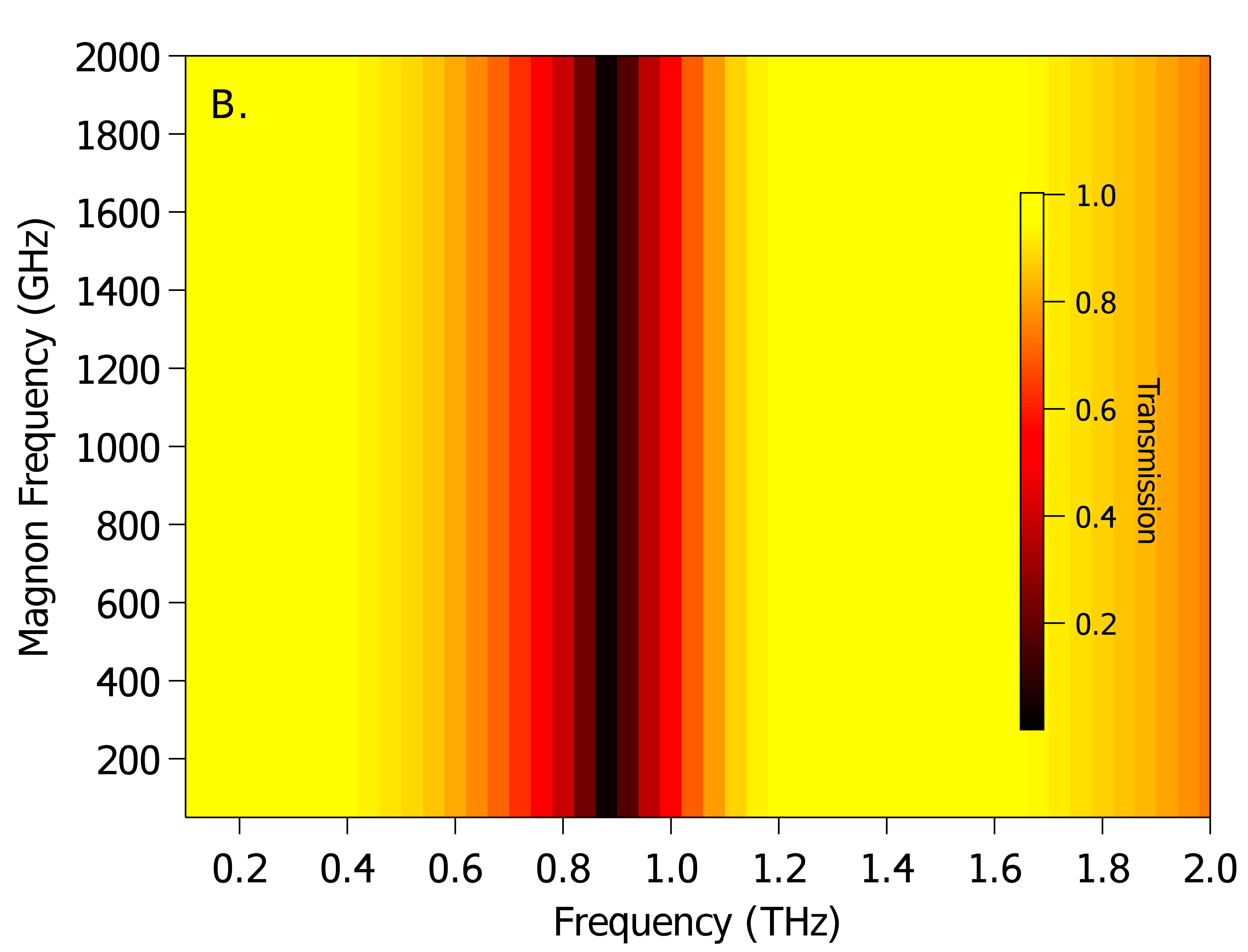}
    \centering
    \includegraphics[width=0.48\columnwidth]{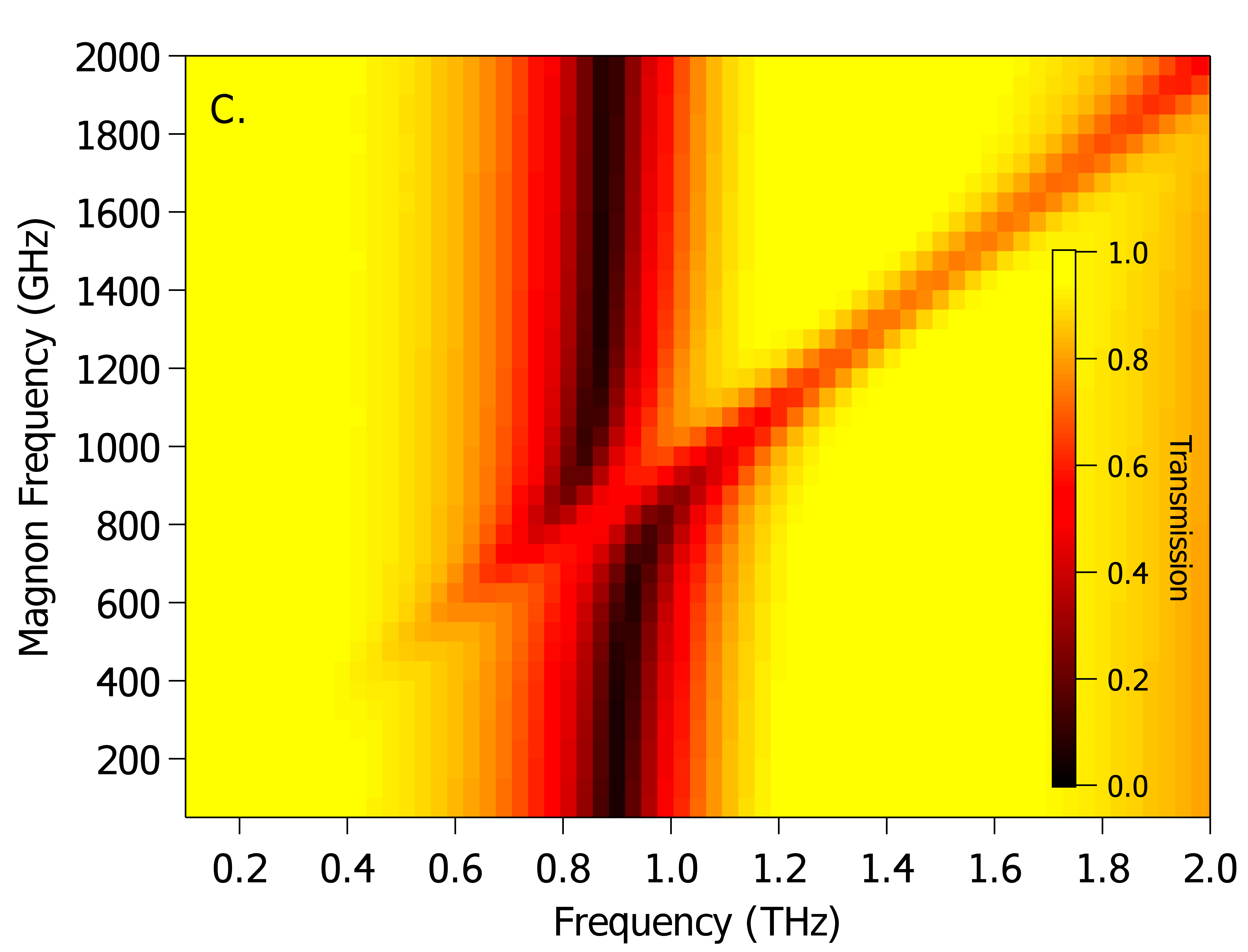}
    \centering
    \includegraphics[width=0.48\columnwidth]{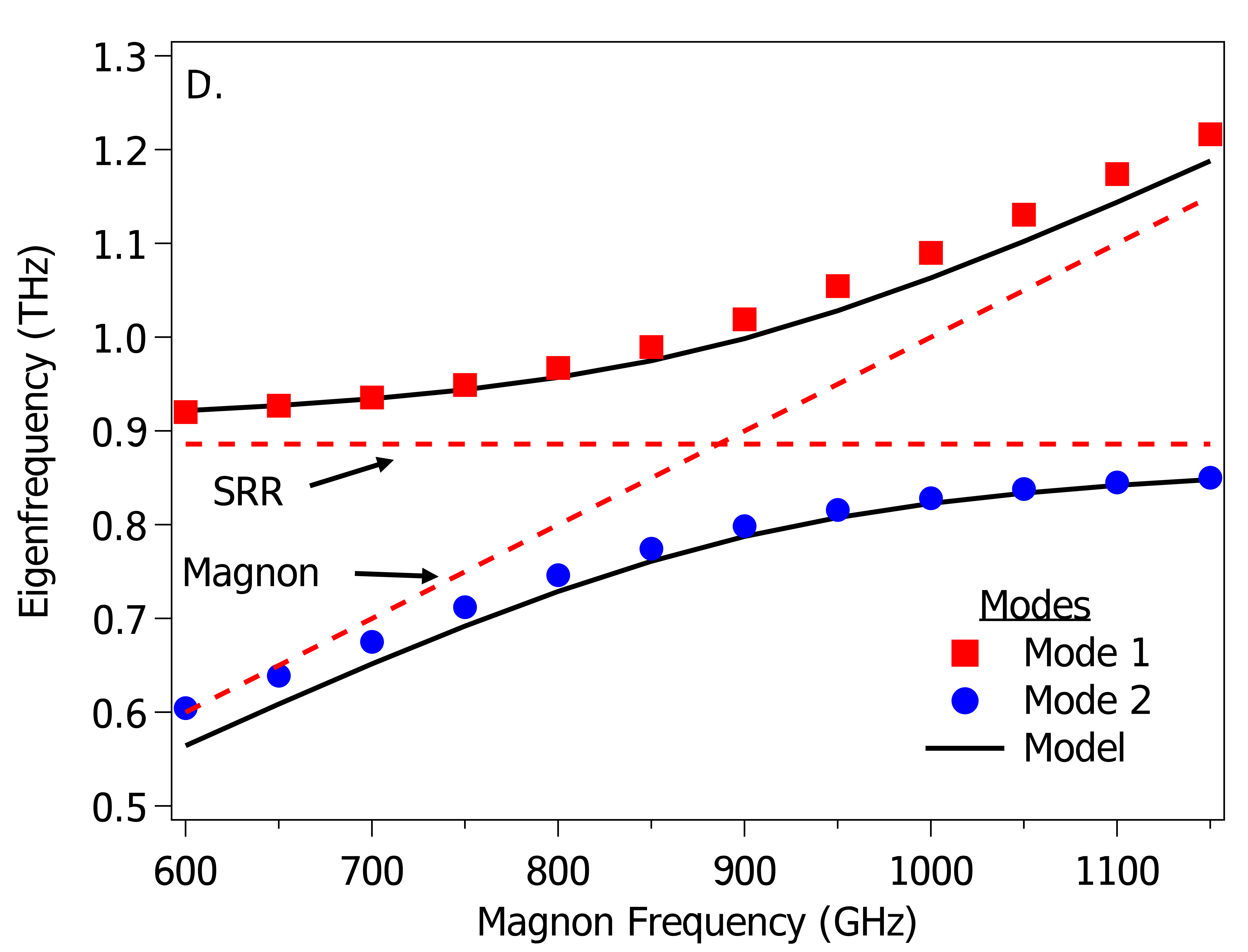}
    \end{center}
    \caption{\textbf{A.} The transmission spectrum of the AFM without the presence of SRRs. Since the spins are oriented parallel to the H-field, the magnon is not excited and will not show up in the signal. An array of SRRs, whose frequency is fixed for all magnon frequencies measured as seen in \textbf{B.}, was then added to the system. \textbf{C.} is the result of this combined system. The magnon mode, although absent in \textbf{A.}, is present and interacts with the SRRs LC mode in the form of a Rabi splitting. The resulting hybridization of the LC/magnon modes due to this splitting is observed in \textbf{D.} Solid lines are from model described in text, dashed lines are the bare LC and magnon resonance.
    }
    \label{fig:hybrid_modes_p50}
\end{figure*}

\section{Results}

\begin{figure}[t!]
    \centering
    \includegraphics[width=.5\columnwidth]{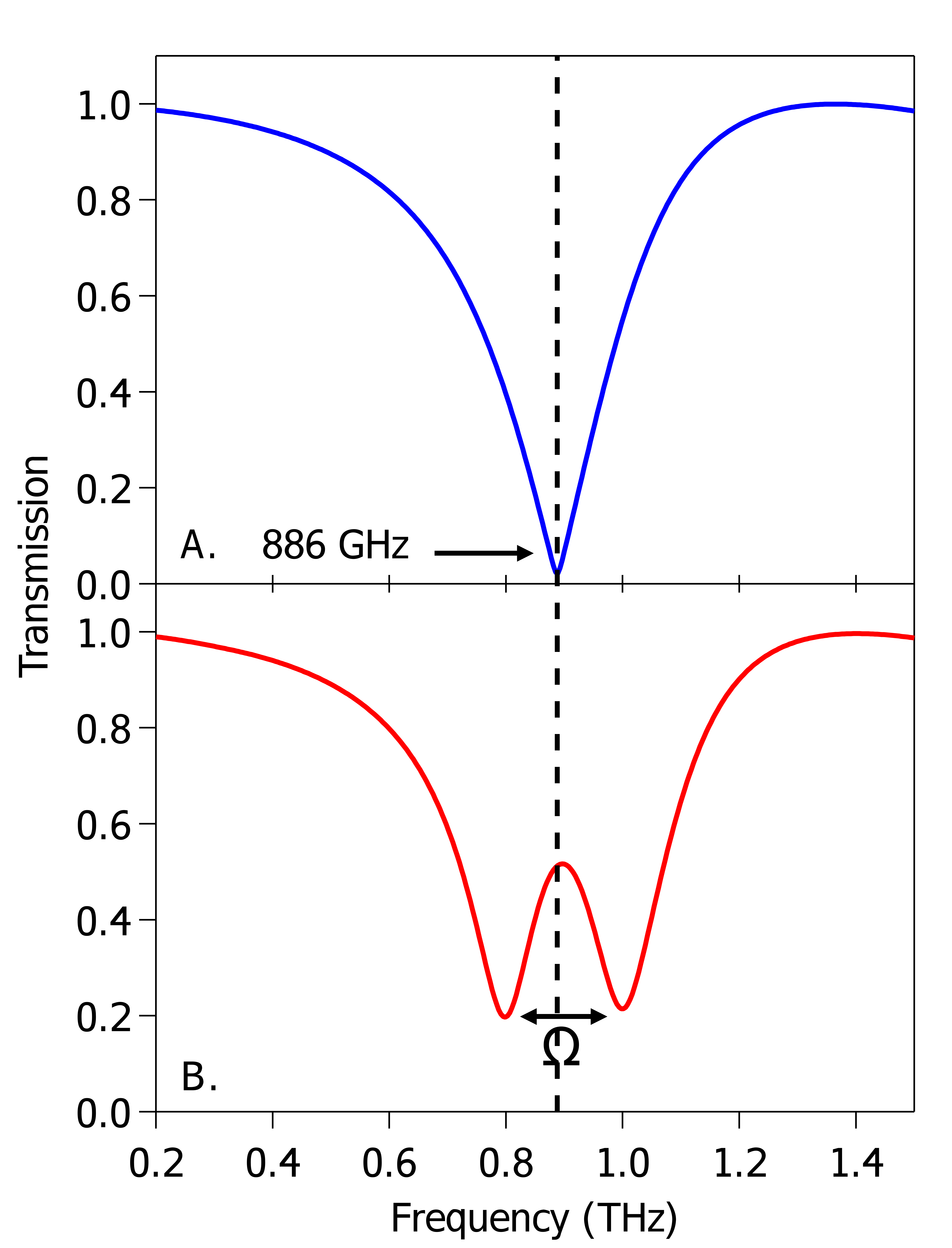}
    %\caption{\textcolor{blue}{This figure is not referred to at all in the text!!}
    \caption{\textbf{A.} Metamaterial LC resonance response simulated on the magnetic material with a non-dispersive permeability of 1 and a permittivity of 6. For the specified metamaterial dimensions, we found the resonant frequency $\sim$886 GHz. \textbf{B.} Transmission when the magnetic material's resonant frequency is equal to that of the metamaterial. The result is an anti-crossing signaled by the Rabi frequency, $\Omega$.}
    \label{fig:c_vs_d}
\end{figure}

To best visualize the coupling, the simulation corresponding to an oscillator strength, $\sigma/\sigma_0$, of 50 was analyzed (Fig. \ref{fig:hybrid_modes_p50}). Three cases are presented: 1) when SRR's are absent (Fig. \ref{fig:hybrid_modes_p50}A.), 2) when only the SRR's are present (Fig. \ref{fig:hybrid_modes_p50}B), and 3) when both the SRR and AFM are present (Fig. \ref{fig:hybrid_modes_p50}C). When SRRs are absent, the AFM showed no response from the applied THz pulse as expected of an AFM with permeability equal to 1. The pure SRRs response is also as expected \cite{o2002magnetic} as the LC resonance responds to the applied THz field, however when both the SRRs and the AFM are present, the transmission shows a clear anti-crossing behavior between two distinct modes (see Fig. \ref{fig:c_vs_d}). At low magnon frequency (i.e. when the magnon frequency is lower than the LC resonance frequency), one mode is present: the LC resonance of the SRR. As the magnon frequency is increased, the magnon hybrid mode begins to emerge. When the magnon frequency is near the LC resonance, an anti-crossing is observed, where the magnon hybrid mode begins to take the role of the LC resonance hybrid mode and the other becomes the new magnon hybrid mode. For this particular value of AFM oscillator strength, the coupling constant, $V$, was found to be $\sim$0.45 meV. This corresponding value for $V$ was used to construct a model using the solutions to the coupled oscillator equation, as seen in Fig. \ref{fig:hybrid_modes_p50}D. The result is in good agreement with the simulations. Surprisingly, at high frequency, the magnon hybrid mode continues to be detected even though the geometry of the experiment would dictate that, as the permeability of the AFM is independent of frequency, no magnon mode should be detected. The source of this effect could be linked to residual off-resonance dipole modes of the SRR. The existence of this mode implies that the magnon mode can be studied for an orientation that would not normally allow for a magnetic response. Furthermore, we expect that the penetration depth of the SRR fields will be limited to the interface region between the metamaterial and the AFM, and thus magnons will only be excited near the surface.

\begin{figure}[!t]
    \centering
    \includegraphics[width=0.7\columnwidth]{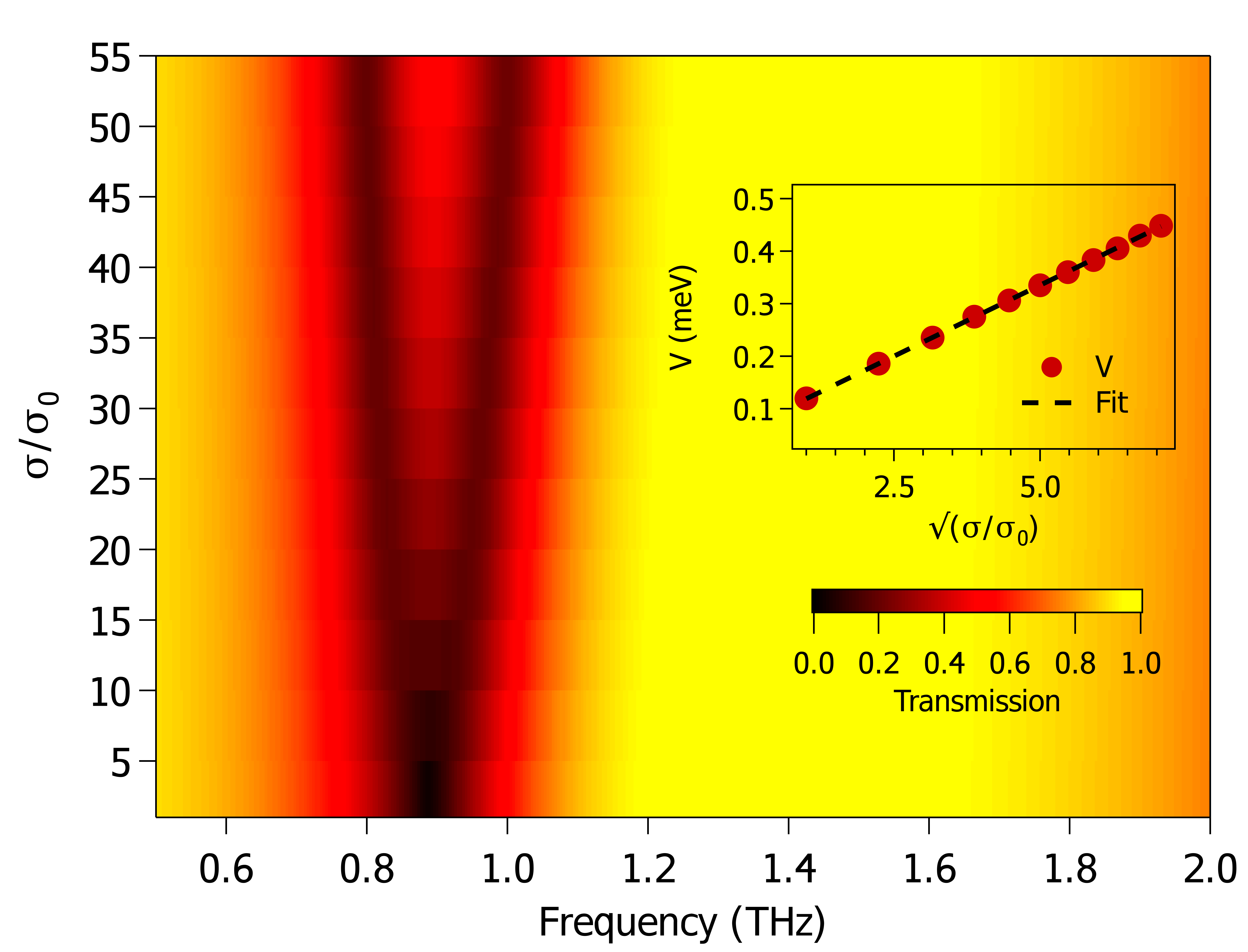}
    \caption[]{Dependence of the transmission on the oscillator strength of AFM, where $\sigma/\sigma_0$ ranges between 1 and 50 at increments of 5. At low $\sigma/\sigma_0$, Rabi splitting is nonexistent as $V$ is in the weak coupling regime. As $\sigma/\sigma_0$ is increased, a separation between modes becomes apparent. The inset shows the calculated $V$ which shows a square root dependence on $\sigma/\sigma_0$. Dashed line is a fit to eq. \ref{eq:sqrt}, where $y$ is 0.065 meV and $A$ is 0.054 meV.}
    \label{fig:oscvscoupling}
\end{figure}

To study the effects at LC-resonance, the strength of the coupling is determined using the two-coupled oscillator model, which relates the behavior of the two hybrid modes to the coupling constant, $V$ \cite{kavokin2017microcavities},
\begin{equation}
    (\omega_{SRR}-\omega-i\gamma_{SRR})(\omega_{AFM}-\omega-i\gamma_{AFM}) = V^2,
    \label{eq:coupling_eq}
\end{equation}
where $\omega_{SRR,AFM}$ and $\gamma_{SRR,AFM}$ are the bare frequencies and the damping coefficients of the SRR/AFM respectively. When solved, $\omega$ is found to have solutions $\omega_{1,2}-i\gamma_{1,2}$, which correspond to the hybrid modes of the coupled oscillator. The constant, $V$, in eq. \ref{eq:coupling_eq} can be determined by using the value of the forbidden energy gap, or Rabi frequency, $\Omega$, between the hybrid modes when the two decoupled oscillators are frequency matched. $\omega_{SRR}$ was obtained by simulating the transmission of the SRRs on AFM, using a dispersionless permeability of 1. It was found that $\omega_{SRR}$ is $\sim$886 GHz with a damping coefficient of $\sim$114 GHz as shown in figure \ref{fig:c_vs_d}A. The magnon frequency, $\omega_{AFM}=2\pi f_0$, was set to this value and the Rabi frequency $\Omega$ was determined from the frequency splitting in the simulation as shown in figure \ref{fig:c_vs_d}B. The coupling constant $V$ was obtained from \cite{kavokin2017microcavities}:
\begin{equation}
    \Omega = \sqrt{4V^2-(\gamma_{SRR}-\gamma_{AFM})^2}.
    \label{eq:coupling}
\end{equation}
These calculations are only valid when $V$ is in the strong coupling regime, 
\begin{equation}
    V>\Big|\frac{\gamma_{SRR}-\gamma_{AFM}}{2}\Big|,
    \label{eq:strong_coupling}
\end{equation}
below which, $\Omega$ is 0 and the resulting $V$ from calculation will remain constant, and it will be equal to $|\gamma_{SRR}-\gamma_{AFM}|$.

The coupling dependence on the magnon oscillator strength of the AFM was studied first. At the LC resonance of the SRR, an increase in coupling with increasing value of $\sigma/\sigma_0$ was observed as expected (Fig. \ref{fig:oscvscoupling}). At low oscillator strength, the separation between hybrid modes is nonexistent. As $\sigma/\sigma_0$ was increased, the transmission minimum increased until a clear separation between two modes became apparent. $V$ was calculated and is shown in the inset of Fig. \ref{fig:oscvscoupling}. $V$ shows a square root dependence on the oscillator strength. This data was fit to a function of the form,
\begin{equation}
    V = y + A \sqrt{\sigma/\sigma_0},
    \label{eq:sqrt}
\end{equation}
where $y$ is 0.065 meV and $A$ is 0.054 meV.

\begin{figure}[!t]
    \centering
    \includegraphics[width=0.7\columnwidth]{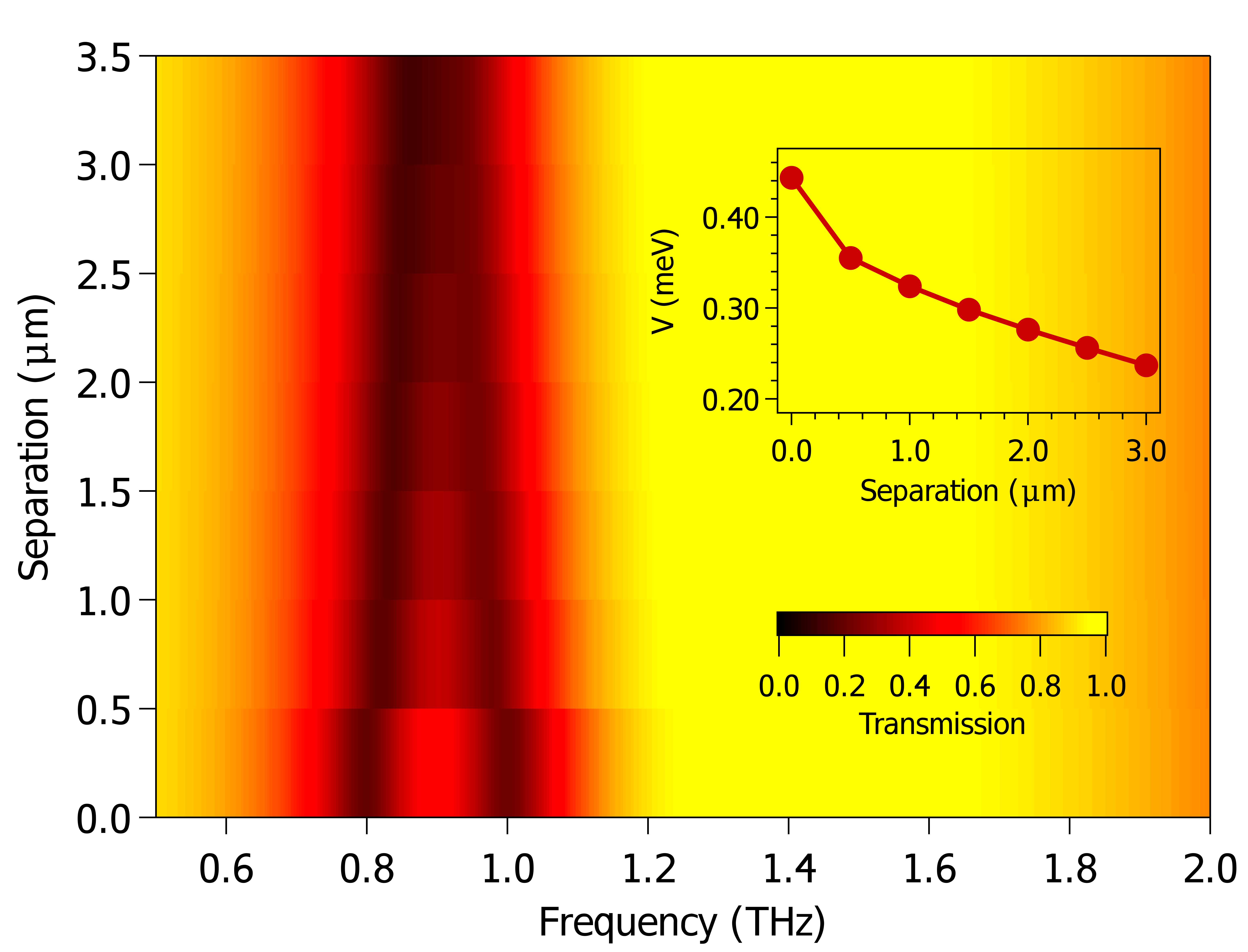}
    \caption[]{The dependence of the transmission on the separation between the SRRs and AFM, where the thickness of the dielectric spacer ranges between 0 $\mu$m and 3 $\mu$m at 0.5 $\mu$m increments. At no separation, a Rabi splitting is observed. With the introduction of a dielectric spacer between the SRRs and AFM, a dramatic decrease in mode separation occurs. As the distance increases, the Rabi splitting continues to decrease until the weak coupling regime is reached. This behavior is shown by $V$ as seen in the inset.}
    \label{fig:coupling_p50}
\end{figure}

To evaluate the extent at which the coupling seeps into the AFM, the separation between the SRRs and AFM was adjusted (Fig. \ref{fig:coupling_p50}), where we used a value of $\sigma/\sigma_0 = 50$. When separation is 0 $\mu$m, the modes are distinguishable from one another, but with the introduction of the dielectric spacer, the Rabi frequency monotonically decreases until the gap becomes nonexistent. This behavior is represented by the derived coupling constants for the various separations (inset of Fig. \ref{fig:coupling_p50}). Here the coupling decreases continuously until it reaches a value that corresponds to when the system enters the weak coupling regime which occurs around 3 $\mu$m.

\section{Discussion}

\begin{figure}[!t]
    \centering
    \includegraphics[width=0.5\columnwidth]{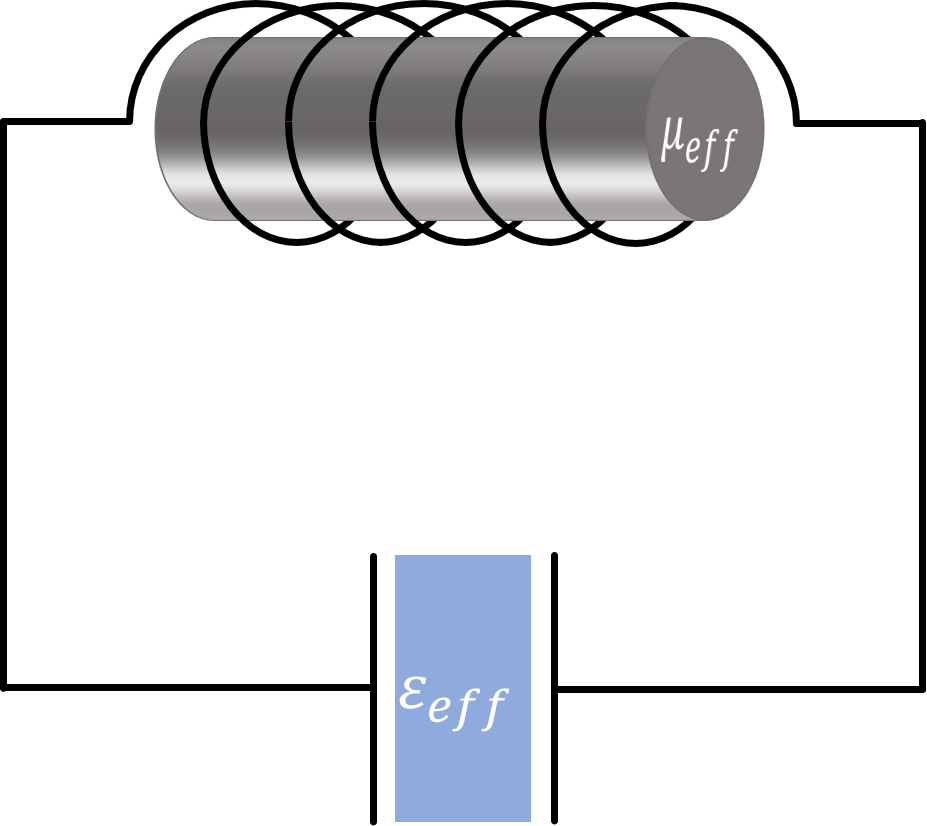}
    \caption{Circuit representation of the SRR/AFM system at the low frequency limit. The coupling between the SRR and AFM can be visualized as an inductor with a magnetic core, whose permeability matches that of the AFM.}
    \label{fig:circuit}
\end{figure}

To understand the source of the coupling, one must first look at the structure chosen for this experiment; the split ring resonator. This structure, at low frequencies can be visualized as an LC circuit \cite{baena2005equivalent}. This consists of an inductor and capacitor in series (Fig. \ref{fig:circuit}). For the polarization chosen for this simulation, the capacitor component is driven by the external electric field from the THz pulse. This in turn generates a surface current that produces a magnetic flux through the inductor component of the SRR, i.e. through the ring portion of the structure. These fields permeate through the substrate, which for this experiment, is an antiferromagnetic material. This will stimulate the magnetic excitations of the AFM corresponding to the $\mu_{xx}$ component of the permeability tensor, hence, generating a magnon response. This "dark mode" associated with AFM influences the "light mode" associated with the SRR and will arise as a feature in transmission signature of the SRR in the form of a Fano resonance.

One can go deeper with this lumped circuit representation to describe the inner workings of this coupling process. The AFM can be assumed to be a magnetic core of the inductor of this circuit. The inductance, in this case, would take the form $L_{eff}=\mu_r(\omega) L$. Thus, the LC circuit, including loss a loss term, $R$, can be represented by
\begin{equation}
    \frac{d^2I}{dt^2}+\frac{R}{\mu_r(\omega)L}\frac{dI}{dt}+\frac{I}{\mu_r(\omega)LC}=0,
    \label{eq:circuit}
\end{equation}
with a solution of the form $I(t) = I_0\exp{(i\omega t)}$. There are three different regimes associated with eq. \ref{eq:circuit}: where $\omega_{AFM} \approx \omega_{SRR}$, $\omega_{AFM} \gg \omega_{SRR}$, and $\omega_{AFM} \ll \omega_{SRR}$.

For the first regime eq. \ref{eq:circuit} can be approximated to:
\begin{equation}
    (\omega_{AFM}-\omega-i\gamma_{AFM})(\omega_{SRR}-\omega-i\gamma_{SRR})=\sigma\omega^2_{AFM}.
    \label{eq:circuit_coupling}
\end{equation}
Setting $V=\sqrt{\sigma}\omega_{AFM}$, eq. \ref{eq:circuit_coupling} takes the form of a coupled oscillator as defined by eq. \ref{eq:coupling_eq}. Noting that $V \propto \sqrt{\sigma}$, one can see a similar relationship between oscillator strength and $V$ by the inset of Fig. \ref{fig:oscvscoupling}. For the two other regimes, the two hybrid modes can be evaluated separately. The LC resonance would become 
\begin{equation}
    \omega_{SRR} = \frac{1}{\sqrt{\mu_r(\omega')LC}},
    \label{eq:SRR_freq}
\end{equation}
where $\omega'\rightarrow\infty$ or $0$ for the low and high magnon frequency limits, respectively. In regards to the magnon hybrid mode at low magnon frequency, the inductive component of the LC circuit representation is negligible compared to the capacitive component of the circuit. Thus, the inductive component would not influence the current of the LC circuit at this magnon frequency limit. At high frequency, the capacitive component becomes negligible and the inductive component dominates. The overall impedance of the LC circuit would then be dependent on the permeability of the magnetic core of the inductor. When the resonant frequency of this model circuit reaches the magnon frequency, the sudden change in the permeability will result in an overall increase in the inductance, whose lineshape would reflect that of the magnon response.

\section{Summary}

The coupling between the SRRs and AFM was numerically evaluated to understand the interaction between the LC resonance of an SRR and a magnon in an AFM material. It was shown that coupling manifested itself in the form of an anti-crossing of the excitations as well as the emergence of a magnon hybrid mode outside of the mode crossing. The coupling showed strong dependence on oscillator strength and the vicinity of the two materials. Materials that show a weak magnetic response, i.e. peak imaginary component of permeability on the order of 0.01, will not show significant coupling. The coupling was found to be mostly concentrated at the surface of the magnetic material. These results provide additional motivation to experimentally explore the effects of magnon-photon couplings using metematerials as an optical cavity.

\section*{Acknowledgements}
We thank Yufei Li and Chase Lyon for comments on the manuscript.

\paragraph{Funding information}
We acknowledge funding from the Center for Emergent Materials, an NSF MRSEC, under grant DMR-2011876. Additional support was provided by OSU's Institute for Materials Research under grants IMR-FG0168 and EMR-G00030.

\bibliography{coupling}

% TODO:
% Provide your bibliography here. You have two options:

% FIRST OPTION - write your entries here directly, following the example below, including Author(s), Title, Journal Ref. with year in parentheses at the end, followed by the DOI number.
%\begin{thebibliography}{99}
%\bibitem{1931_Bethe_ZP_71} H. A. Bethe, {\it Zur Theorie der Metalle. i. Eigenwerte und Eigenfunktionen der linearen Atomkette}, Zeit. f{\"u}r Phys. {\bf 71}, 205 (1931), \doi{10.1007\%2FBF01341708}.
%\bibitem{arXiv:1108.2700} P. Ginsparg, {\it It was twenty years ago today... }, \url{http://arxiv.org/abs/1108.2700}.
%\end{thebibliography}

% \bibliographystyle{SciPost_bibstyle} % Include this style file here only if you are not using our template
%\bibliography{SciPost_Example_BiBTeX_File.bib}

\nolinenumbers

\end{document}